\documentstyle[12pt,aaspp4]{article}

\lefthead{Molteni, T\'oth, \& Kuznetsov}
\righthead{Azimuthal Stability of Accretion Shocks}

\begin{document}

\title{On the Azimuthal Stability of Shock Waves around Black Holes}

\author{Diego Molteni}
\affil{
Dipartimento di Scienze Fisiche ed Astronomiche Universit\`a di Palermo
Via Archirafi 36, 90123 Palermo, Italy} 
%molteni@gifco.fisica.unipa.it

\author{G\'abor T\'oth}
\affil{Department of Atomic Physics, E\"otv\"os University,
P\'azm\'any s\'et\'any 2, Budapest 1117, Hungary}
%gtoth@hermes.elte.hu, http://hermes.elte.hu/\~{}gtoth

\and

\author{Oleg A. Kuznetsov}
\affil{Keldysh Institute of Applied Mathematics
Miusskaya sq. 4, 125047 Moscow, Russia} 
%kuznecov@spp.Keldysh.ru

\begin{abstract}
Analytical studies and numerical simulations of time dependent
axially symmetric flows onto black holes have shown that it is possible to
produce stationary shock waves with a stable position both for ideal
inviscid and for moderately viscous accretion disks.

We perform several two dimensional numerical simulations of accretion flows
in the equatorial plane to study shock stability against {\em 
non-axisymmetric azimuthal perturbations}. We find a peculiar new result. A
very small perturbation seems to produce an instability as it crosses the
shock, but after some small oscillations, the shock wave suddenly transforms
into an asymmetric closed pattern, and it stabilizes with a finite radial
extent, despite the inflow and outflow boundary conditions are perfectly
symmetric.

The main characteristics of the final flow are: 1) The deformed shock
rotates steadily without any damping. It is a permanent feature and the
thermal energy content and the emitted energy vary periodically with time.
2) This behavior is also stable against further perturbations. 3) The
average shock is still very strong and well defined, and its average radial
distance is somewhat larger than that of the original axially symmetric 
circular shock.
4) Shocks obtained with larger angular momentum exhibit more frequencies and
beating phenomena. 5) The oscillations occur in a wide range of parameters,
so this new effect may have relevant observational consequences, like
(quasi) periodic oscillations, for the accretion of matter onto black holes.
Typical time scales for the periods are 0.01 and 1000 seconds for black
holes with 10 and $10^6$ solar mass, respectively.
\end{abstract}

%%%%%%%%%%%%%%%%%%%%%%%%%%%%%%%%%%%%%

\keywords{accretion, accretion disks ---  instability --- shock waves}

%%%%%%%%%%%%%%%%%%%%%%%%%%%%%%%%%%%%%

\section{Introduction}

Shock waves in rotating accretion flows onto compact objects present a very
important mechanism to transform the potential gravitational energy into
radiation. The possibility that they exist also around black holes (BH) has
been suggested years ago by Hawley, Smarr, \& Wilson (\markcite{hsw84a}
1984a, \markcite{hsw84b}1984b). However their shocks were unstable in radial
direction. Recently it has been shown that there is a wide range of
parameters (namely specific angular momentum and initial temperature) for
which steady shock configurations are possible (Chakrabarti \markcite{c89}
1989, \markcite{c90}1990; for viscous flows see Chakrabarti \& Molteni
\markcite{cm93}1993, Lanzafame et al. \markcite{lmc98}1998). Essentially the
shock position depends on thermal energy and angular momentum contents of
the flow: the balance between the centrifugal and total pressure forces (ram
plus internal) is the main factor determining the shock formation. The
stability was examined with axially symmetric numerical calculations
(Molteni, Lanzafame, \& Chakrabarti \markcite{mlc94}1994; Chakrabarti \&
Molteni \markcite{cm95}1995). Recent analytical studies (Lu \& Yuang
\markcite{ly97}1997; Nakayama \markcite{n93}1993, \markcite{n94}1994) have
also shown the stability of isothermal and adiabatic shocks against axially
symmetric perturbations.

The stability of shock waves in accretion flows has great astrophysical
relevance since, in this way, the emission due to the shock is not a
transient episode, but can be a permanent mechanism responsible of the
radiated energy. However, the studies made up to now are all in axial
symmetry.

In this paper we examine the azimuthal stability of the initially 
axisymmetric shocks by perturbing them with an {\em asymmetric perturbation}.
We explore the parameter space defined by the angular momentum and
temperature of the accreted gas. The computations are done in the equatorial
plane assuming slab symmetry in the $z$ direction (thin disk approximation).
We use two different computer codes to integrate numerically the
hydrodynamic equations: the Versatile Advection Code (VAC, T\'oth
\markcite{t96}1996, \markcite{t97}1997, also see
http://www.phys.uu.nl/\~{}toth/) and another one developed by Kuznetsov
(Kuznetsov et. al. \markcite{k98}1998).

Here we are not discussing the origin of the supersonic accretion flow
and the influence of viscosity. The formation of supersonic flows
in the accretion disk has been predicted by various authors in different 
contexts valid both for galactic and extragalactic supermassive Black Holes 
(for viscous transonic flows cfr. Chakrabarti \markcite{c96}1996, 
for advection dominated flows see Gammie and Popham \markcite{gp98}1998, 
Igumenshev, Abramowicz, \& Novikov \markcite{ian98}1998, 
Narayan, Mahadevan, \& Quataert \markcite{naq98}1998).
These works clearly show that since the viscous time scale is much longer
than the infall time scale close to the black hole, the angular momentum
remains almost constant, unless the viscosity is really very high (i.e. 
the $\alpha$ parameter close to unity). 
The supersonic region is well inside the accretion disk 
(typically within 50 Schwarzschild radii), 
far from the impact of any accretion stream that
typically occurs at $10^6$ Schwarzschild radii.
Therefore the supersonic inflow is expected to be 
approximately axisymmetric. Small non-axisymmetric
perturbations may occur, of course, which is exactly the motivation 
for our study.

The paper is structured as follows. Sections 2 and 3 describe the analytic
and numerical models, respectively, the results of the numerical simulations
are presented in Section 4, and conclusions with our plans for further
investigations are discussed in Section 5.

%%%%%%%%%%%%%%%%%%%%%%%%%%%%%%%%%%%%%

\section{Modeling the problem}

We solve numerically the hydrodynamic equations describing the time
dependent motion of a rotating, inviscid, ideal gas falling into a
Schwarzschild black hole.
We assume an initial axisymmetric configuration
and a permanent axisymmetric boundary condition inflow. Since our study
concerns the inner regions of the accretion disk ($ r \le 50 r_g $) we may
disregard the possible perturbation due to spiral shocks (Yukawa et al. 
\markcite{ybm97}1997).
Indeed, the validity of the extension of their result interior into the disk 
is far to be clear.
Furthermore, the phenomenon, we discuss, may occur in accretion onto black 
holes in galactic nuclei, where a fully axisymmetric boundary condition 
may easily occur.
To allow for analytical comparison and to be able
to study the phenomenon with reasonable computer time, the motion is
restricted to the $z=0$ plane and we assume that gradients in the $z$
direction are negligible, i.e. $\partial /\partial z=0$. 
The general relativistic effects are taken into account by using the
Paczy\'nski \& Wiita \markcite{pw80}(1980) pseudo-Newtonian gravitational
potential
\begin{equation}
\label{q-pot}\Phi (r)=-{\frac{GM}{r-r_g}}=-\frac{c^2}{2(r/r_g-1)}\,,
\end{equation}
where $r,G,c,M$ and $r_g$ are the radial distance, the gravitational
constant, the speed of light, and the mass and the Schwarzschild radius 
($r_g=2GM/c^2)$ of the accreting object, respectively. This approximation is
frequently used when accurate relativistic details are not required. 
The results show that both the initial axisymmetric and the final
distorted shocks form at radial distances ($r \geq 5 r_g$) 
for which the pseudo-Newtonian potential is fairly accurate. 

We solve the equations of hydrodynamics in polar coordinates $r$ and 
$\varphi$ for mass density $\rho$, radial momentum density $\rho v_r$,
angular momentum density $r \rho v_\varphi$ and energy (kinetic$+$thermal)
density $e $. The four equations express conservation of mass, radial
momentum balance, conservation of angular momentum, and energy balance,
respectively.
\begin{eqnarray}
  \frac{\partial r\rho}{\partial t}
  +\frac{\partial r\rho v_r}{\partial r}
  +\frac{\partial \rho v_\varphi}{\partial \varphi}&=&0
                                                         \label{q-rho} \\
  \frac{\partial r\rho v_r}{\partial t}
  +\frac{\partial }{\partial r      }(r\rho v_r^2+r p)
  +\frac{\partial }{\partial \varphi}(\rho v_r v_\varphi)
  &=&\rho v_\varphi^2 + p - r\rho\frac{\partial \Phi}{\partial r}
                                                        \label{q-momr}\\
  \frac{\partial r^2\rho v_\varphi}{\partial t}
  +\frac{\partial}{\partial r     }(r^2\rho v_r v_\varphi)
  +\frac{\partial}{\partial\varphi}(r\rho v_\varphi^2+r p)
  &=&0
                                                       \label{q-angmom} \\
  \frac{\partial r e}{\partial t}
  +\frac{\partial}{\partial r}[r v_r(e+p)]
  +\frac{\partial}{\partial \varphi }[v_\varphi (e+p)]
  &=&-r\rho v_r\frac{\partial \Phi }{\partial r}
                                                        \label{q-e}
\end{eqnarray}
where the thermal pressure is $p=(\gamma-1)[e-\rho(v_r^2+v_\varphi ^2)/2]$
with the polytropic index $\gamma=4/3$, which is appropriate for
relativistic gas. We adimensionalize our equations by using the
Schwarzschild radius and the speed of light as reference units, i.e. $r_g=1$
and $c=1$.

The outer boundary of the computational domain is set far away from the BH
at a radius $r_{max}$ (typically $r_{max}\sim 20 - 50 r_g$), but closer than
the outer sonic point, so that the inflow is supersonic. The inner boundary
is chosen at a distance $r_{min}$ inside the inner sonic point where the
flow behind the shock becomes supersonic again due to the gravitational
acceleration. Typically we take $r_{min}\sim 1.5 r_g$. The boundary
conditions are periodic for the $0 \le \varphi < 2\pi$ coordinate.

Since the equations are linear in the mass and energy densities, we may
choose the density of the infalling gas at the outer boundary of the
computational domain to be unity, $\rho(r_{max})=1$. Once the density scale
is fixed, the steady state solution is fully determined by three conserved
quantities: the accretion rate $A=r\rho v_r$, the specific angular momentum 
$\lambda=r v_\varphi$ and the Bernoulli constant $B=\Phi+(e+p)/\rho$
(sometimes referred to as `energy'). To match the full solution, which
contains an outer sonic point outside the domain, the accretion rate cannot
be chosen independently of the other two constants, rather it should be
calculated from the set of the algebraic equations expressing the mass
conservation, the energy conservation and the polytropic equation of state
following the procedure outlined in the Appendix.
The remaining two parameters, the angular momentum and the Bernoulli
constant can be chosen such that the steady shock is positioned at a
distance $r_{shock}$. The dependence of the shock position on $\lambda$ and 
$B$ is shown in Fig.~\ref{f-anal} in the parameter range of our test cases.
The supersonic inflow parameters are independent of the azimuthal coordinate
$\varphi$ and, if not perturbed, the initial steady state flow would remain
axi-symmetric forever.

\placefigure{f-anal}

The initial condition is the steady solution perturbed a few units upstream
of the shock at $r_{perturb}\sim r_{shock}+3r_g$ and $\varphi=\pi$ in a
small region of size $\Delta r=r_{perturb} \Delta\varphi=0.05 r_{perturb}$.
We typically perturb the pressure or the density by a few per cent only,
therefore the energy and momentum content of the perturbation is completely
negligible relative to the respective quantities for the full domain. The
perturbation is advected into the BH with the flow as shown by the upper
panels of Fig.~\ref{f-movie}. 

\placefigure{f-movie}

%%%%%%%%%%%%%%%%%%%%%%%%%%%%%%%%%%%%%%%%%%%%%%%%%%%%%%%

\section{Numerical scheme}

To integrate numerically the hydrodynamic equations (\ref{q-rho})-(\ref{q-e}),
we use two different programs, VAC (T\'oth \markcite{t96}1996,
\markcite{t97}1997) and another code implemented by Kuznetsov (Kuznetsov et.
al. \markcite{k98}1998). Both codes use Total Variation Diminishing (TVD,
Harten \markcite{h83}1983) type schemes. In VAC we used a characteristic
based Lax-Wendroff type TVD scheme with a Roe type approximate Riemann
solver (Roe \markcite{r86}1986) and minmod or Woodward flux limiters (see
T\'oth \& Odstr\v cil \markcite{to96}1996 for details), while in the other
code a Lax-Friedrichs type scheme is applied with Osher-Chakravarthy
anti-diffusion term (Chakravarthy \& Osher \markcite{co85}1985). The
combined Lax-Friedrichs-Osher scheme was suggested by Vyaznikov, Tishkin, \&
Favorsky\markcite {vtf89}(1989). Here the $r$ and $\varphi$ fluxes are added
at the same time, while VAC uses a Strang type dimensional splitting. Both
schemes are 2nd order accurate in space, the VAC code is temporally second
order too, while the other code is only first order accurate in time.
There are further differences in the grid settings and in the implementation
of many numerical details. We have checked for several cases that the two
codes give essentially identical results.

The analytical equations are discretized in conservation form. In both
codes, the physical and numerical fluxes are calculated for the angular
momentum density $r\rho v_\varphi$ instead of the usual tangential momentum
density $\rho v_\varphi$ to achieve better angular momentum conservation. In
VAC, this does not require the modification of the approximate Riemann
solver itself, only the obtained fluxes are handled differently.

Another concern is the temporally accurate evaluation of source terms on the
right hand sides of the equations (\ref{q-momr}) and (\ref{q-e}), which
arise from the curvature of the coordinate system and the gravitational force
of the BH. The Lax-Friedrichs-Osher code is only first order accurate in
time, thus very small time steps are used (Courant number $\approx 0.1$). In
VAC, we follow the algorithm described by Ryu et. al. \markcite{r95}(1995)
in their Appendix with a correction of their equation (A22), which should
read $\bar q^{n+1/2}=(q^n+q^n_{hydro}+S^n)/2$, to achieve second order
accuracy. This allows larger time steps with Courant numbers $\approx0.8$.
A naive implementation of a Lax-Wendroff type TVD scheme in polar
coordinates leads to first order temporal accuracy and very inaccurate
steady state shock positions for Courant numbers close to unity.

Boundary conditions are easily implemented with two layers of ghost cells
surrounding the physical domain. Supersonic inflow and outflow can be
realized with {\it fixed} and {\it extrapolated} ghost cell values,
respectively. The fixed cell values are derived from the flow parameters:
the accretion rate $A$, angular momentum $\lambda$, and the Bernoulli
constant $B$, which are listed for all the test cases in Table 1.
Periodicity in the $\varphi$ direction is trivial to maintain.

Convergence of the numerical results was checked by redoing the same
calculation with different grid resolutions $N_r\times N_{\varphi} =
200\times60, 200\times120, 200\times100$, and $300\times90$. We also have
the option of using a logarithmic grid spacing in the $r$ direction
(typically we use $\Delta r_{j+1}/\Delta r_j\approx1.003$), which allows a
better resolution of sharp gradients close to the BH. The results were found
to be extremely similar on all grids. The evolution becomes qualitatively
wrong only for a very coarse grid spacing $100\times30$, where the
oscillations are suppressed by numerical diffusion and the perturbed shock
returns to the symmetric steady state after a short transient. We may
conclude that the results are not too sensitive to a small amount of
dissipation.

Although an analytical steady state can be easily derived by solving
ordinary differential equations (Chakrabarti \& Molteni 
\markcite{cm93}1993), 
it is better to obtain a {\it numerical equilibrium}, which is always
slightly different from the analytical steady state solution due to the
discretization errors. In particular, VAC can solve the time dependent
axially symmetric one dimensional equations starting from a crude initial
condition. The most efficient way to converge to steady state is to do a 1D
simulation with axial symmetry employing a fully implicit time stepping
algorithm (see T\'oth, Keppens, \& Botchev \markcite{tkb98}1998 for details).

The 1D equilibrium solution is `rotated' around the symmetry axis to get the
starting configuration for the 2D runs, i.e. the 2D variables 
$w^{(2D)}(i_r,i_\varphi)=w^{(1D)}(i_r)$ for $w=\rho,\rho v_r, r\rho
v_\varphi, e$, where $i_r,i_\varphi$ are the grid indices. The radial grid
spacing must be the same for the 2D and 1D grids, but the number of grid
points in the $\varphi$ direction can be chosen freely. We have tested that
{\it without} the non-axisymmetric perturbation the 2D steady state obtained
this way is stable and no fluctuations are introduced in the flow, since the
numerical scheme used for the 2D simulation is the same as for the 1D
simulation.

Once a steady solution with the shock is obtained, the small perturbation
described at the end of the previous section is added in the cells around 
$r=r_{perturb}$, $\varphi=\pi$ and the flow is evolved with the 2D code using
explicit time stepping for hundred thousands of time steps.

%%%%%%%%%%%%%%%%%%%%%%%%%%%%%%%%%%%%%%%%%%%%%%%%%

\section{Results}

Several test cases were studied in the parameter space defined by the
specific angular momentum $\lambda=r v_\varphi$ and the Bernoulli constant 
$B=\Phi+(e+p)/\rho$. Beside $\lambda$ and $B$, Table 1 also lists the scaled
accretion rate $A/\rho(r_{max})=-r v_r \rho/\rho(r_{max})$, which is
necessary for setting up the boundary conditions, and the shock position 
$r_{shock}$ for the axially symmetric steady state, which can be obtained by
the procedure described in the Appendix. The cases are ordered by the radial
shock distance and the specific angular momentum: the bigger case numbers
(1, 2, ... 5) correspond to larger $r_{shock}$, and the subcases (a, b, ...)
have larger and larger $\lambda$, respectively. The outer boundary is located
at $r_{max}=56,\ 20,\ 25,\ 50$ and $56$ for the five cases respectively.

\placetable{table}

To check for time variations, we monitor the flow for the maximum shock
radial distance, for the shock position at $\varphi=0$, and for the mass and
thermal energy contents of different sectors and annuli. We also analyzed
snapshots at different times (see Fig.~\ref{f-movie}), 
and visualized the simulation with animations.

All cases, except the stable case 4, evolve rather similarly. As the
perturbation crosses the hot subsonic postshock region, the shock becomes
slightly distorted, and it oscillates with a small amplitude. After this
transient, the phases of the small oscillations synchronize, and a very clear
distorted shock develops with the end closed back as shown by 
Fig.~\ref{f-shockfig}. Further details of the final shock structure
can be studied on Fig.~\ref{f-shockfig}.

\placefigure{f-shockfig}

\placefigure{f-pvr}

The new axially asymmetric shock, with a dominant $m=1$ mode, reaches a new
``quasi steady state'' with finite radial extent (in case 5b the simulation
had to be stopped when the shock touched the outer boundary of the
computational domain). The radial distance range of the twisted shock is not
very different from the axially symmetric shock position $r_{shock}$. The
quasi steady state means that the flow changes periodically, without any
sign of damping or instability. In case of ``regular oscillations'' the flow
pattern rotates like a solid body with a period $P$, so it is a true steady
state in the co-rotating coordinate system. In other cases more than one
frequencies are present, which means that the solid body rotation is
superimposed with a genuine oscillation of the flow pattern. When the ratio
of the two periods are close to a small integer, ``beating'' can be observed
in the monitored quantities. When many periods are present, the time
variations become irregular, although the amplitude remains bounded. Table 1
summarizes the results and the observed periods for all cases.

Certain trends can be easily identified from the table. For a fixed shock
distance $r_{shock}$, the larger the specific angular momentum $\lambda$ is,
the more and usually longer periods are observed. This can also be seen in
Fig.~\ref{f-mass1}, where the time variation of the total mass in a angular
sector $0<\varphi<2\pi/3$ is plotted. The figure contains further
information. Note that the higher $\lambda$ is, the larger the average mass
in the sector becomes, i.e. the further away the deformed shock is from the
BH, since most of the mass is contained in the compressed post shock region.
Furthermore, the amplitude of the mass variation also increases with 
$\lambda $, which corresponds to a larger radial extent of the asymmetric
shock pattern. Even in the most unstable case 1d, however, the total mass in
the sector increases by about $30\%$ only.

\placefigure{f-mass1}

A clean example of the ``beating'' phenomenon can be seen in Fig.~\ref
{f-mass2c}, which shows the total mass variation in the same angular sector
for case 2c. The Fourier spectrum of the time variation, obtained with a
discrete FFT, shows two well defined peaks with almost identical amplitudes
in Fig.~\ref{f-fourier2c}. Since the ratio of the frequencies is close to 2,
strong beating can be observed in the time variation.

\placefigure{f-mass2c} \placefigure{f-fourier2c}

The periods $P$ are in the same range $60<P<280$ for all cases listed in
Table 1, despite the large variation in the rotation period at the shock
distance, which is $2\pi r^2_{shock}/\lambda$ and it varies roughly from 100
to 2000 for cases 1 to 5, respectively. The radial extent of the deformed
shock is not too large in most cases, e.g. for cases 2a and 3 the radial
shock distance varies between 9 to 11 and 13 to 16, respectively. In case
5b, however, the shock has reached the outer boundary, at $r_{max}=56$,
which is more than twice the shock distance $r_{shock}=23.4$ of the axially
symmetric steady state. 

%%%%%%%%%%%%%%%%%%%%%%%%%%%%%%%%%%%%%%%%%%%%%%%%%%%%%

\section{Conclusions}

We find that the axisymmetric shocks predicted by Chakrabarti's theory
(Chakrabarti \markcite{c90}1990) in a rotating inviscid accretion flow are
generally unstable to azimuthal perturbations, but {\em the instability
saturates at a low level, and a new, stable, asymmetric configuration
develops with a strongly deformed shock rotating steadily}. It is possible
that shocks produced with very large angular momentum are unstable, in the
sense that the perturbation triggers large deviations from the circular
symmetry that extend up to the outer sonic point, but this does not happen
in the test cases presented here.

The new disk is no longer axisymmetric despite that the boundary conditions
are the same as initially: axially symmetric supersonic inflow at the outer
and supersonic outflow at the inner boundary. According to the numerical
simulations, the asymmetric shock configuration is continuously self
sustained and self reproducing around the BH. In general, this asymmetric
configuration of the flow speed, density, and temperature produces time
variations in any measured quantities. For a fixed shock distance we find
that the time variations are regular periodic for low angular momentum flows
and they are more irregular, containing more frequencies, for flows with a
larger angular momentum.

The time variation could be observed and it could be a signature of the
black hole's presence. As it can be seen in Table 1, typical periods are in
the range 60 to 300 in the dimensionless units. Taking $P=100$ as a typical
case, we may convert it to physical units by multiplying with 
$2GM/c^3\approx 10^{-5}M/M_\odot\,$sec. For a black hole of ten solar mass,
this gives approximately 0.01 seconds, while a BH with $M=10^6M_\odot$ would
produce oscillations typically with 1000 second periods.

We point out that although the general scenario seems different, there are
similarities with the nonaxisymmetric disk instabilities studied by Blaes \&
Hawley \markcite{bh88}(1988). Of course other important physical ingredients
have to be included for a more realistic study of the shock behavior: full
3-dimensional treatment, physical viscosity, true cooling mechanism etc.
Fully three dimensional simulations using the
Versatile Advection Code of the same problem are in due course and will be
presented in a forthcoming
work. The role of the physical viscosity should also be
further investigated, since in general, as shown in a similar context (cfr.
Lanzafame, Molteni, \& Chakrabarti \markcite{lmc98}1998), but in
axisymmetric conditions, moderate viscosity may produce changes in the shock
structure and induce oscillations, while a large viscosity may stabilize the
flow in a Keplerian regime, eliminating the shocked solutions. Our
convergence studies indicate that the azimuthal instability is not very
sensitive to a small amount of ``numerical viscosity''. In any case we
note that it has
already been shown by time independent studies that the inner part of canonical
accretion disks may have supersonic flows even for the viscosity parameter 
$\alpha$ about 0.01 
(Chakrabarti, \markcite{c96}1996) or even larger 
(Igumenshev, Abramowicz, \& Novikov \markcite{ian98}1998).

Our investigation is intended as a first exploration of the new axially
asymmetric solutions. Despite the many simplifying assumptions, we suggest
that such a phenomenon may occur easily in real physical systems when
viscosity is small and initial angular momentum is sub-Keplerian since the
parameter range that leads to shocked solutions is fairly large and the
perturbation required to trigger the asymmetric configuration is very small.
The azimuthal instability described in this paper, in any case, is a fine
example of how non linearity of the fluid dynamic equations may break the
symmetry of the initial and boundary conditions.

\acknowledgements
VAC was developed by G.T. as part of the project on `Parallel Computational
Magneto-Fluid Dynamics', funded by the Dutch Scientific Research Foundation
(NWO) Priority Program on Massively Parallel Computing. The first
simulations were done on a Cray C90 for which computer time was sponsored by
the Dutch National Computing Facilities Foundation (NCF). G.T. currently
receives a postdoctoral fellowship (D~25519) from the Hungarian Science
Foundation (OTKA). 
O.A.K. was supported by the Russian Foundation for Basic Research
(grant 97-02-16486). Both G.T. and O.A.K. thank
the University of Palermo for its hospitality during their visits.

\appendix

\section{Algebraic Equations for 1D Steady State Solution}

\newcommand{\MACH}{{\cal M}}

Here we give a method to calculate the 1D steady state solution,
in particular the scaled accretion rate $A/\rho(r_{max})$ and the shock
position $r_{shock}$ as a function of the specific angular momentum $\lambda$
and the Bernoulli constant $B$.
This method requires (iterative) solutions of algebraic equations
without the need to integrate ordinary differential equations
(cfr. Chakrabarti \markcite{c89}1989, \markcite{c90}1990 for a method
using integration).

Let us introduce the sound speed $a=(\gamma p/\rho)^{1/2}$ and
the Mach number in the radial direction $\MACH=-v_r/a$.
In case of an axially symmetric one dimensional steady state, i.e.
$\partial/\partial t=\partial/\partial\varphi = \partial/\partial z=0$,
we can integrate (\ref{q-rho}), (\ref{q-angmom}), and (\ref{q-e})
to obtain three conserved quantities, $A=r \rho v_r$,
$\lambda=r v_\varphi$, and $B=(e+p)/\rho+\Phi$, respectively.
For any continuous isentropic solution, the density and the sound speed
are related as $\rho=Ca^{2/(\gamma-1)}$, where $C$ is a constant
for the particular solution. We may now proceed to eliminate
$\rho$, $v_r$, $v_\phi$, $p$, and $e$ in favor of the conserved quantities
$A$, $B$, $C$, and $\lambda$, and the single unknown function,
the Mach number $\MACH(r)$. First $a^2$ should be expressed from the
Bernoulli equation
\begin{equation}\label{q-bern}
  B=  \frac 12 (\MACH a)^2+\frac{\lambda ^2}{2r^2}+
  \frac{a^2}{\gamma -1}-\frac{1}{2(r-1)}
\end{equation}
then it can be substituted into the mass conservation equation
\begin{equation}
  A= - r \rho v_r = C r \MACH a^{\frac{\gamma+1}{\gamma-1}}
\end{equation}
to arrive at the final equation for $M(r)$
\begin{equation}\label{q-fg}
  \frac A C = f(\MACH) \cdot g_{\lambda,B}(r)
\end{equation}
with
\begin{equation}
  f(\MACH) =\frac \MACH{\left[ \frac 12\MACH^2+\frac{1}{(\gamma -1)}\right]
        ^{\frac{\gamma+1}{2( \gamma -1) }}}
\end{equation}
and
\begin{equation}
  g_{\lambda,B}(r) =r\cdot
   \left[ B-\frac{\lambda ^2}{2r^2}+\frac{1}{2(r-1)}\right]
   ^{\frac{\gamma +1}{2(\gamma -1) }}
\end{equation}
The $f$ function has a single maximum at $\MACH=1$, and it can be inverted
both in the subsonic $0\le \MACH<1$ and supersonic $\MACH>1$ regions.
The $g$ function
has in general two local minima at $r_1$ and $r_2$ with $r_1>r_2$,
which can be determined by solving the algebraic equation $dg/dr=0$
numerically.

At large distances from the BH, the Mach number is $\MACH\ll 1$, while
close to the horizon $\MACH\gg 1$, thus any continuous solution has
to have a sonic point with $\MACH=1$, where $f[\MACH(r)]$ has a maximum.
Since the $f(\MACH)\cdot g(r)$ product must be constant along the flow
(\ref{q-fg}),
the maximum of $f$ should be at one of the minima of $g$, i.e. at
$r_1$ or $r_2$. Therefore we can have two isentropic solutions $\MACH_1$ and
$\MACH_2$
\begin{equation}\label{q-mach}
 \MACH_{1,2}(r)=f^{-1}\left[\frac{f(1)g_{\lambda,B}(r_{1,2})}{g_{\lambda,B}(r)}
                    \right]
\end{equation}
where we use the subsonic branch of $f^{-1}$ for $r<r_{1,2}$ and the
supersonic branch for $r>r_{1,2}\,$.

The outer boundary conditions can now be easily determined from the
$\MACH_1(r)$ solution, which connects the outer sonic point at $r_1$ with
the boundary located at $r_{max}$. First $\MACH_1(r_{max})$ is calculated
from the algebraic
equation (\ref{q-mach}), then the sound speed $a(r_{max})$ from (\ref{q-bern}),
and finally the radial inflow speed $v_r=-\MACH_1 a$ can be obtained.
The scaled accretion rate is $A/\rho(r_{max})=-r_{max}v_r(r_{max})$.

A standing shock can occur in the solution at $r_{shock}$ if
$\MACH_1(r_{shock})>1$ and $\MACH_2(r_{shock})<1$ are related by the Hugoniot
relation
\begin{equation}\label{q-hugoniot}
 \MACH_2=h(\MACH_1)\equiv
  \left[\frac{2+(\gamma-1)\MACH_1^2}{2\gamma \MACH_1^2-(\gamma-1)}\right]^{1/2}
\end{equation}
First $\MACH_1(r_{shock})$ can be determined by solving the algebraic equation
\begin{equation}\label{q-machshock}
  \frac{f(\MACH_1)}{f[h(\MACH_1)]}=
  \frac{g_{\lambda,B}(r_1)}{g_{\lambda,B}(r_2)}
\end{equation}
next the shock position
$r_{shock}=g_{\lambda,B}^{-1}[f(1)g_{\lambda,B}(r_1)/\MACH_1(r_{shock})]$
can be calculated.
In general there can be two solutions, but only the outer one is stable
(Nakayama \markcite{n94}1994).

\newpage

\figcaption[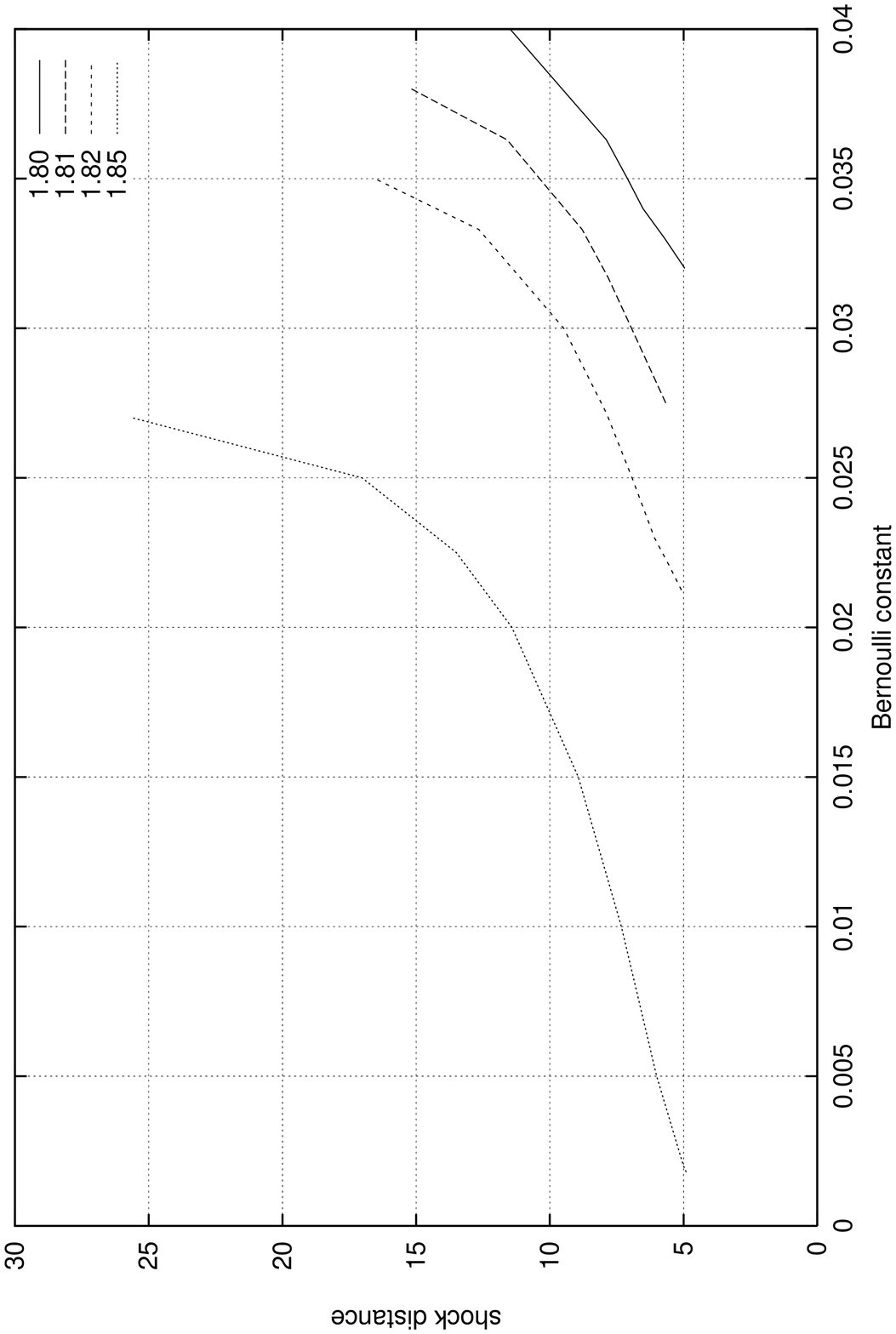]{
  Shock location $r_{shock}$ versus the Bernoulli constant
  $B\equiv(e+p)/\rho+\Phi$
  is plotted for different values of the specific angular momentum
  $\lambda\equiv rv_\varphi$.
\label{f-anal}}

\figcaption[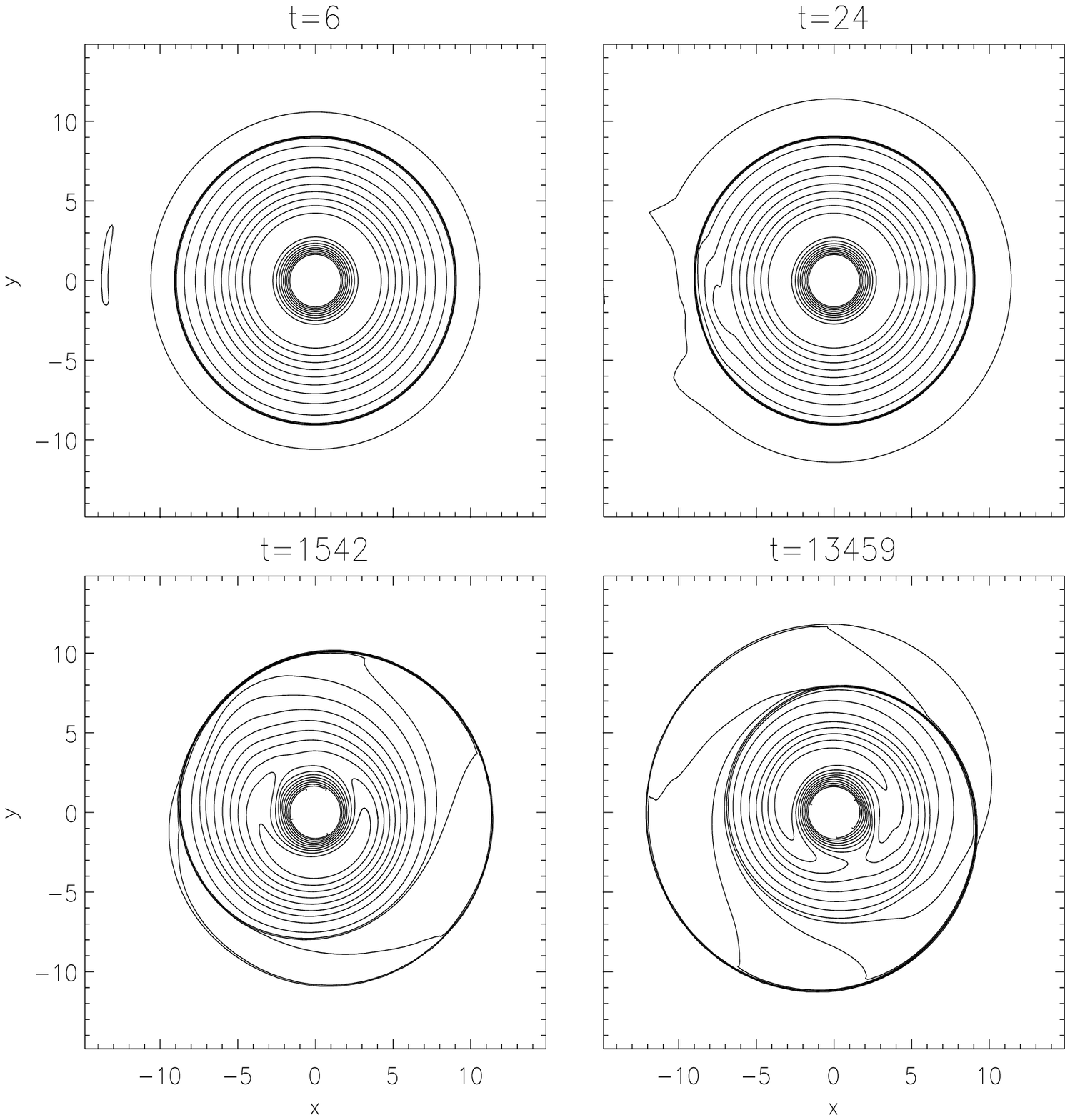]{
  Contourlines of density are shown at four different
  times $t$. The simulation is done in $r, \varphi$ coordinates, but the
  picture is shown in the $x, y$ plane for easier interpretation.
  The initial perturbation has not reached the circular shock in the
  upper left snapshot, and it just crosses the shock front at $t=24$. The new
  twisted shock appears after a long time $t\approx 1000$ and it remains
  stable indefinitely. The parameters of this simulation are similar to
  those of case 2a in Table 1.
\label{f-movie}}

\figcaption[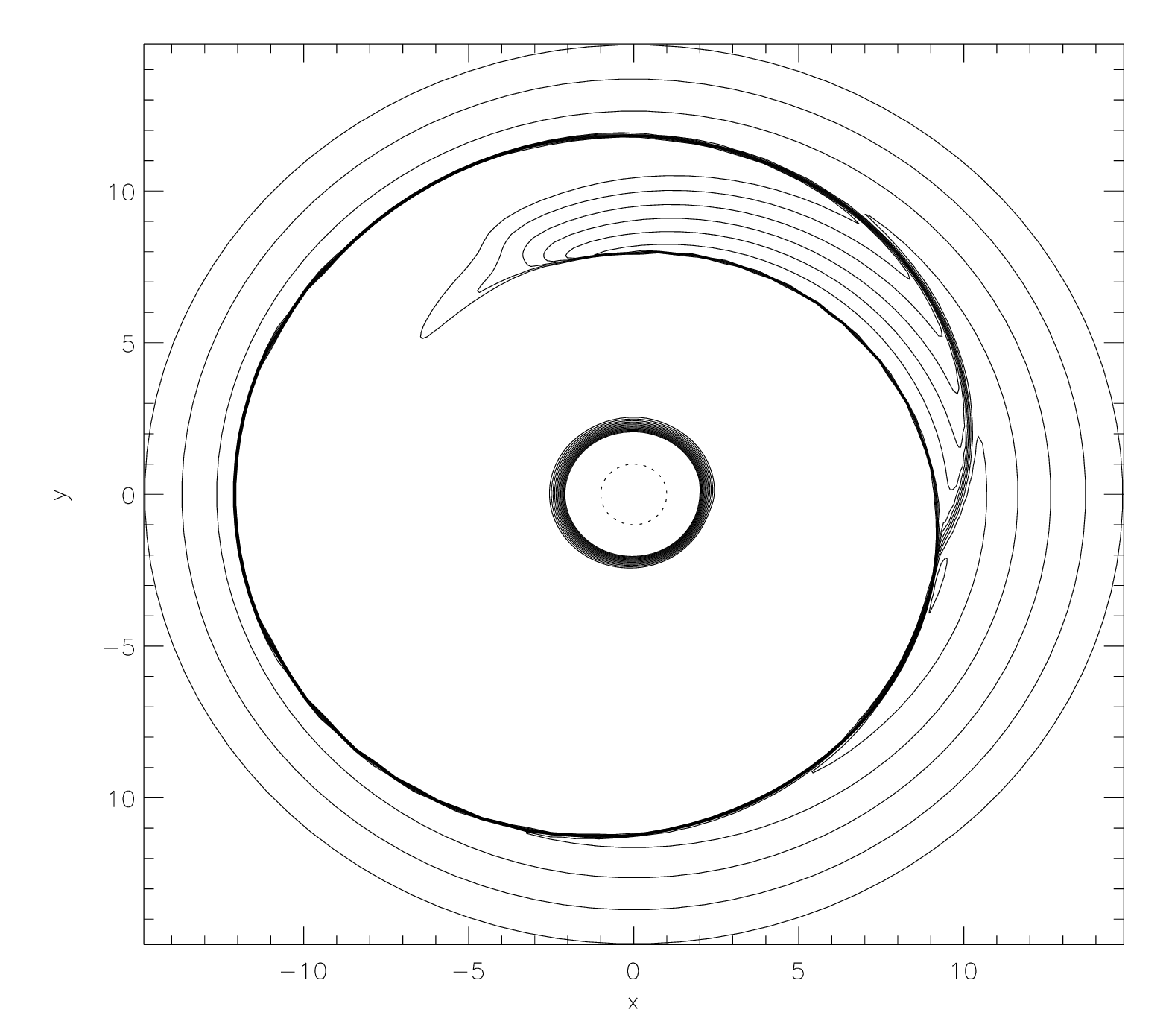]{
  Contour plot of the radial Mach number $|v_r|/a$
  for the last snapshot ($t=13459$) of Fig.~\ref{f-movie}.
  Contourlines in the subsonic region (Mach number below unity) 
  are suppressed so that the shock location is shown as the last 
  contour line of the incoming accretion flow.
  The inner region with unresolved contourlines corresponds to the
  supersonic region of the postshock infall.
  The Schwarzschild radius is indicated with the innermost dotted circle.
\label{f-shockfig}}

\figcaption[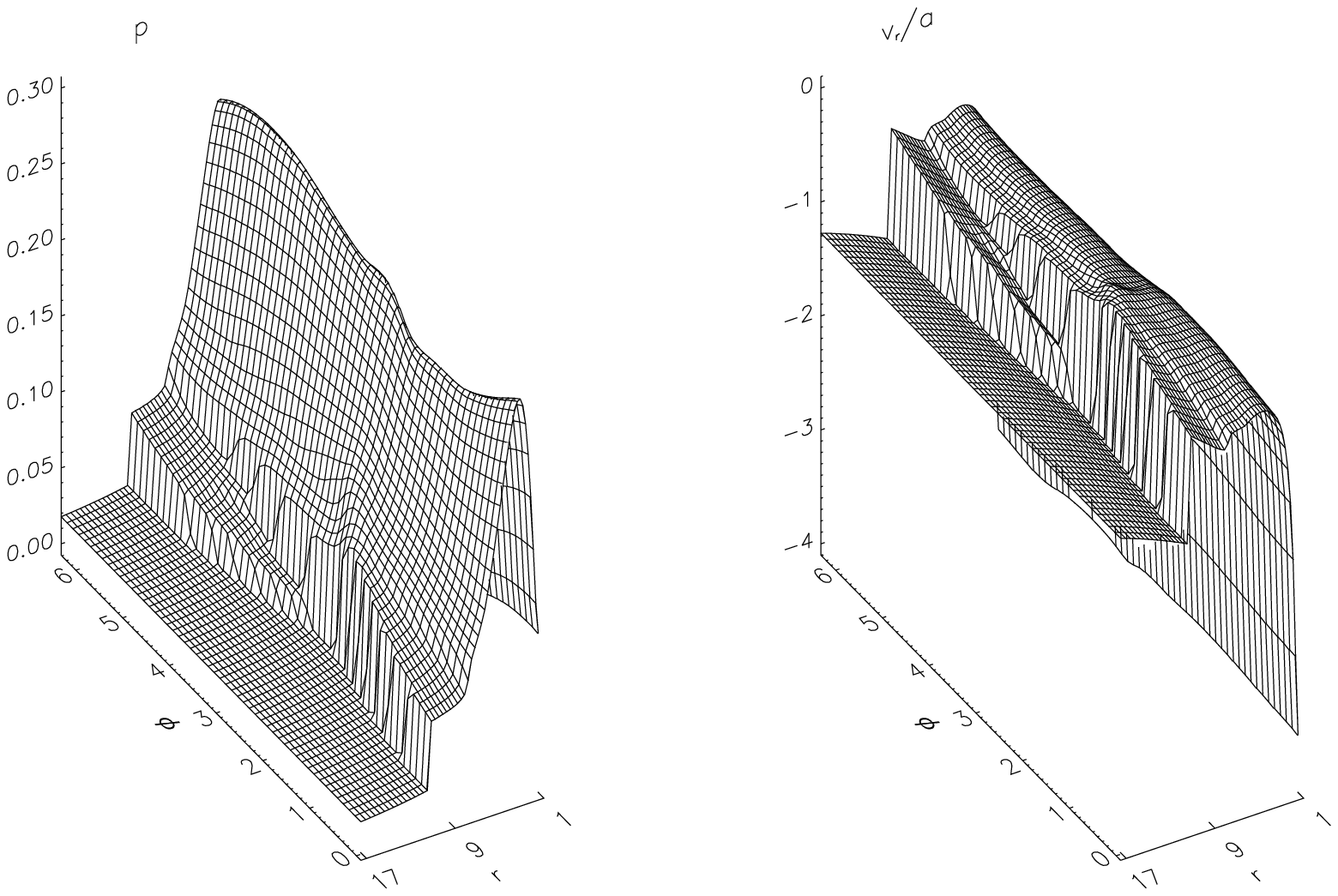]{
  Wire frame representation of pressure $p$ and radial Mach number
  $v_r/a$ in the polar coordinates $r, \varphi$. The black hole is
  at $r=0$ just outside the right boundary.
  For sake of clarity only $50$ grid lines are drawn in both directions,
  but the actual resolution is $200\times100$.
  The shock happens to close back into itself at $\varphi\approx\pi$.
  The density contourlines
  corresponding to this snapshot are shown in
  the left bottom panel of Fig.~\ref{f-movie}.
\label{f-pvr}}

\figcaption[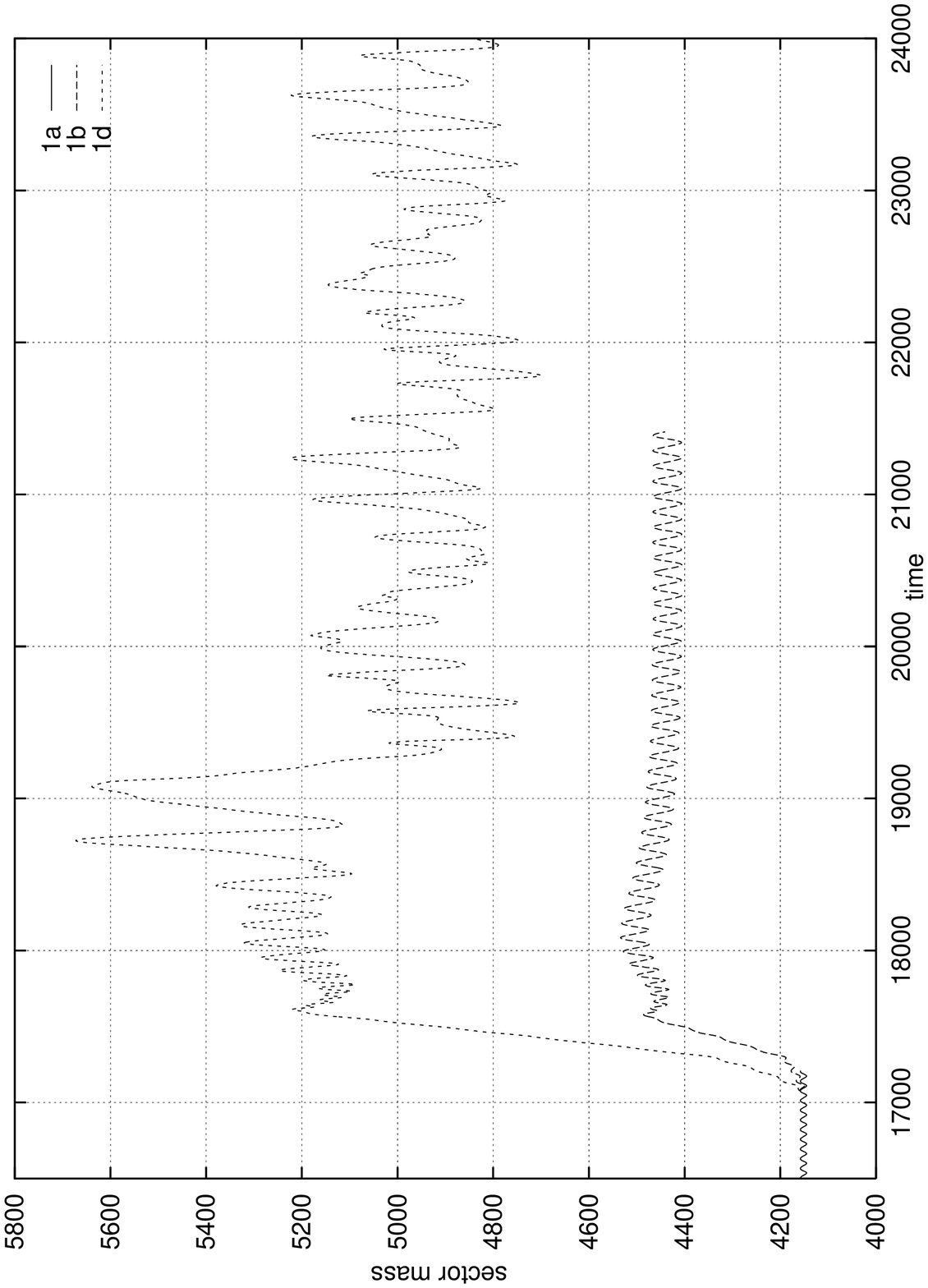]{
  Time variation of the total mass in an angular sector $0<\varphi<2\pi/3$
  for cases 1a, 1b, and 1d, i.e. fixed shock distance $r_{shock}=5.3$
  with increasing specific angular momentum $\lambda$.
  The cases were run until a the amplitudes of the oscillations became
  bounded, thus the curve for case 1a is in the lower left part of the plot.
\label{f-mass1}}

\figcaption[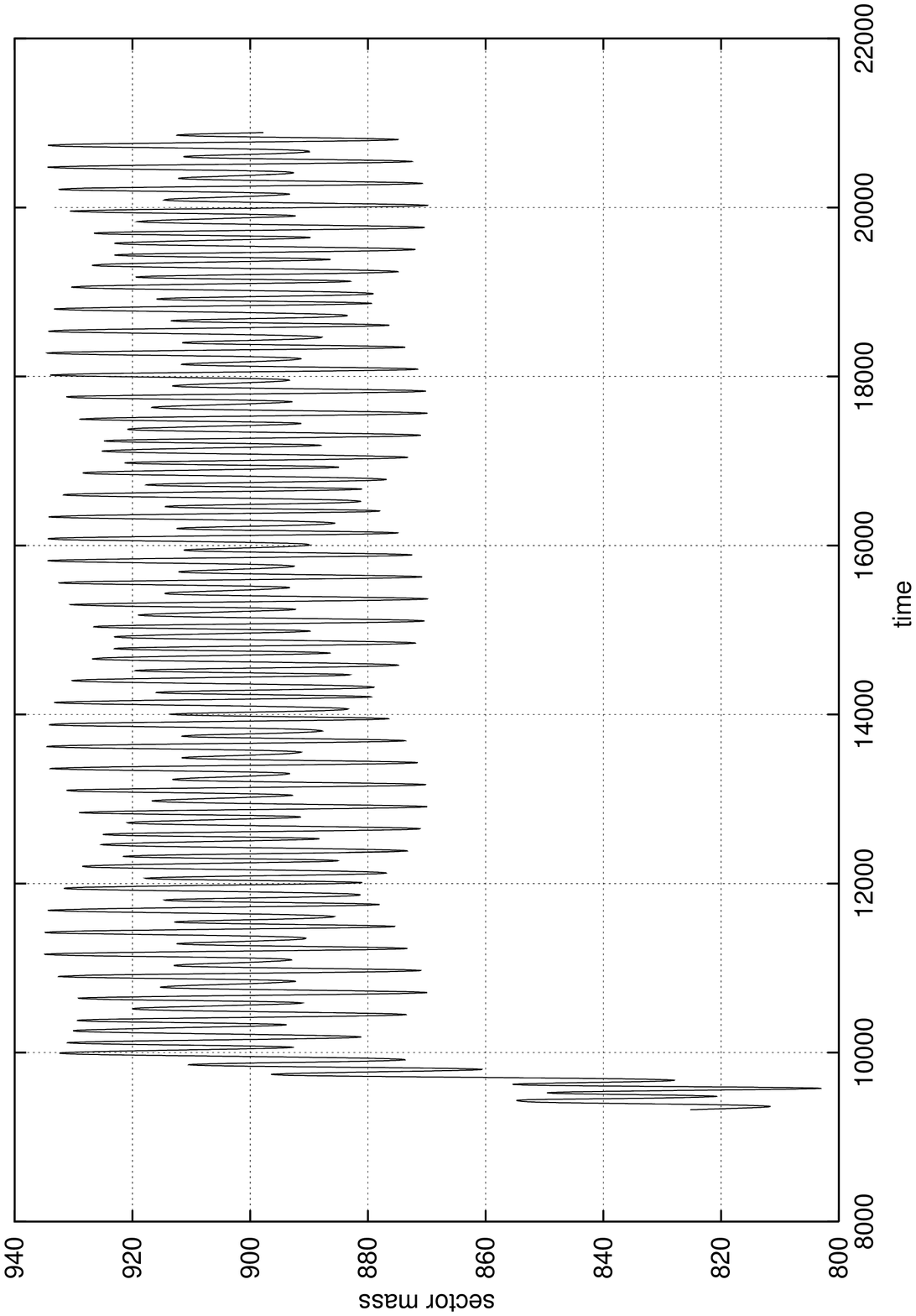]{
  Time variation of the total mass in an angular sector $0<\varphi<2\pi/3$.
\label{f-mass2c}  }

\figcaption[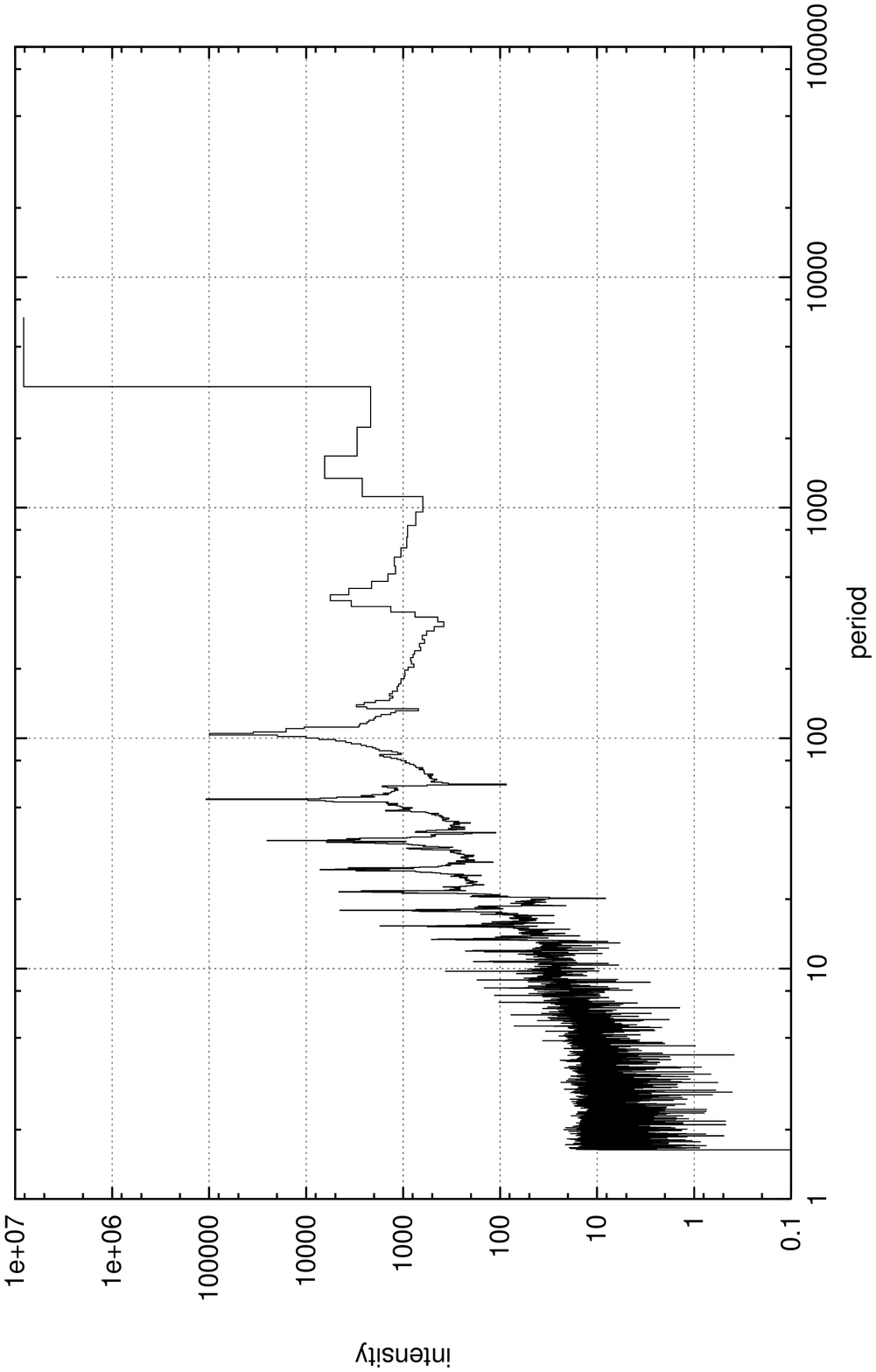]{ \label{f-fourier2c}
  Fourier transform of the time variation shown in Fig.~\ref{f-mass2c}.
  Note the two distinct peaks in the Fourier spectrum at $P=134$ and $P=259$
  with almost equal amplitudes. The closeness of the period ratio to 1:2 is
  responsible for the ``beating''.
}

\newpage

\begin{deluxetable}{lcccccl}
\tablecaption{Simulation Parameters and Results \label{table}}
\tablehead{
\colhead{case}				&
\colhead{$r_{shock}$}			&
\colhead{$A/\rho(r_{max})$}		&
\colhead{$\lambda$}			&
\colhead{$B$}				&
\colhead{$P$}				&
\colhead{comment}}
\startdata
% r_max=56.4
1a& 5.3 & 4.0890 & 1.7770 & .04333 &    63   & regular osc. \nl
1b& 5.3 & 4.4984 & 1.8000 & .03220 &    98   & regular osc. \nl
1c& 5.3 & 4.7262 & 1.8100 & .02701 & 111,209 &  beating     \nl
1d& 5.3 & 5.1041 & 1.8225 & .02000 & \nodata & irregular    \nl
\tablevspace{5pt}
%r_max=20
2a& 7.8 & 3.0000 & 1.8000 & .03630 &  125    & regular osc. \nl
2b& 7.8 & 3.0896 & 1.8100 & .03170 & 242,270 & beating      \nl
2c& 7.8 & 3.1834 & 1.8200 & .02715 & 134,259 & beating      \nl
\tablevspace{5pt}
%r_max=25
3& 12.7 & 3.3450 & 1.8200 & .03332 & 215     & regular osc. \nl
\tablevspace{5pt}
%r_max=50
4& 17.2 & 4.2750 & 1.8255 & .03331 & \nodata  & nearly stable\nl
\tablevspace{5pt}
%r_max=55.86
5a& 23.4 & 4.8766 & 1.8620 & .02307 & 273     & regular osc. \nl
5b& 23.4 & 5.0442 & 1.8720 & .02000 & \nodata  & leaves domain\nl
\enddata
\end{deluxetable}

\end{document}